\documentclass[twocolumn,prd,nofootinbib,aps,prl,floats,floatfix,amsmath,amssymb,longbibliography,secnumarabic]{revtex4-1} %
\usepackage[english]{babel}
\usepackage{csquotes}
\usepackage{csquotes}
\usepackage[dvipsnames]{xcolor}

\usepackage[letterpaper,top=2cm,bottom=2cm,left=3cm,right=3cm,marginparwidth=1.75cm]{geometry}

\usepackage{amsmath}
\usepackage{graphicx,slashed}
\usepackage[colorlinks=true,citecolor=red]{hyperref}

\newcommand{\be}{\begin{equation}}
\newcommand{\ee}{\end{equation}}

\newcommand{\bea}{\begin{eqnarray}}
\newcommand{\eea}{\end{eqnarray}}
\newcommand{\nn}{\nonumber}

\begin{document}

\title{Big Bang Nucleosynthesis Constraint on $B$-mesogenesis}
\author{James M.\ Cline$^{a,d}$}
\email{jcline@physics.mcgill.ca}
\author{Jonas Frerick$^b$}
\author{Sebastian A.\ Naja$^c$}
\author{Yong Xu$^a$}
\affiliation{$^a$ McGill University Department of Physics \& Trottier Space Institute, 3600 Rue University, Montr\'eal, QC, H3A 2T8, Canada}
\affiliation{$^b$ Dipartimento di Fisica, Sapienza Universit\`a di Roma \& Sezione INFN Roma1, Piazzale
Aldo Moro 5, 00185, Roma, Italy}
\affiliation{$^c$ Department of Physics, Faculty of Arts and Sciences, American University of Beirut, Beirut, Lebanon}
\affiliation{$^d$ Niels Bohr International Academy,The Niels Bohr Institute,
Blegdamsvej 17, DK-2100 Copenhagen Ø, Denmark}

\begin{abstract}
$B$-mesogenesis is a low-temperature baryogenesis mechanism, that uses $CP$-violation of the $B$ mesons in the standard model to fulfill the Sakharov conditions. It requires a late period of matter domination by a scalar field $\Phi$ that decays into $b\bar b$ quarks, reheating the Universe below the QCD phase transition.  We investigate the impact of this mechanism on Big Bang Nucleosynthesis, with respect to the reheating temperature $T_{rh}$, and the mass of $\Phi$. 
Injection of hadrons skewing neutron/proton interconversions are the most important effect, followed by neutrino heating from $\Phi$ decays.
For $M_\Phi\lesssim 100\,$GeV, $T_{rh}$ must be $\gtrsim 10\,$MeV. 

\end{abstract}

\maketitle

\section{Introduction} The need for new physics to explain the baryon asymmetry of the Universe is widely recognized.  Although the standard model (SM) of particle physics has anomalous baryon violation at high temperatures and CP violation, necessary ingredients for baryogenesis \cite{Sakharov:1967dj}, attempts to achieve the observed baryon asymmetry within the SM proved to be insufficient
\cite{Farrar:1993sp,Farrar:1993hn,Gavela:1993ts,Gavela:1994dt}.  Electroweak baryogenesis \cite{Cohen:1993nk,Barni:2025ifb} and leptogenesis \cite{Fukugita:1986hr,Davidson:2008bu} are popular mechanisms that use the SM sphalerons to violate baryon number, while introducing new sources of CP violation.  

A complementary approach was proposed more recently, 
called $B$-mesogenesis, that takes advantage of CP violation already present in the $B$ mesons, instead positing new sources of baryon violation \cite{Elor:2018twp,Alonso-Alvarez:2021qfd,Elahi:2021jia}.  It requires a scalar field $\Phi$ that matter-dominates the Universe at late times, which decays to $b+\bar b$ quarks with a low reheating temperature $T_{rh}$, below the QCD phase transition.  The $b$ quarks therefore hadronize into $B$ mesons and undergo CP-violating oscillations before they decay.  New invisible decay channels are hypothesized, that convert the CP asymmetry into the baryon asymmetry.

Previous studies have assumed that the reheat temperature satisfies $T_{rh} > 4\,$MeV to avoid unwanted modifications of the successful predictions of Big Bang Nucleosynthesis (BBN) \cite{Kawasaki:2000en, Hannestad:2004px, deSalas:2015glj}.  The goal of the present work is to better quantify this lower bound.   We start in  Section \ref{RHD} by deriving the nonstandard time-temperature relation that arises during reheating.   Section \ref{NTS} discusses the effects of late reheating and neutrino injection from $\Phi$ decays on the neutrino temperatures.  In Section
\ref{TDBA} we quantify the nonstandard time-dependence of the baryon asymmetry that can arise from late reheating, and show that it has a relatively weak effect on BBN yields for the parameters of interest.
In contrast, injection of energetic mesons by $\Phi$ decays can be important, as discussed in Section \ref{HIE}.  The resulting confidence intervals  on the model parameters are presented in Section
\ref{XRS}, Fig.\ \ref{fig:contours}.  We summarize in Section \ref{SumS}. Throughout this work, we use natural units $\hbar =c=k_B=1$.

\medskip
\section{Reheating dynamics} 
\label{RHD} We start by describing the modified expansion history of the Universe during reheating by $\Phi$ particle decays.
It is well known that the reheat temperature $T_{rh}$ is not the same as the maximum temperature $T_{\rm max}$
\cite{Giudice:2000ex}, whose value depends on initial conditions.  During reheating, the Universe is approximately matter dominated with scale factor $a(t)\sim t^{2/3}$,  but the temperature goes as
$T_{\rm max}\,a^{-3/8}\sim t^{-1/4}$.  It decreases more slowly than usual because radiation is being produced by the decays, while the Universe expands.
When $\rho_\Phi = \rho_{\rm rad}$, most $\Phi$ particles have decayed, and $T$ reaches $T_{rh}$, after which time $T\sim T_{rh}/a \sim t^{-1/2}$, and the conventional radiation domination period proceeds.  If $T_{rh}$ is low enough, the unconventional expansion history preceding $T_{rh}$ can have an impact on BBN.
A common estimate of $T_{rh}$ follows from setting the Hubble rate $H = \sqrt{8\pi G \rho_r/3}$ equal to the $\Phi$ decay rate $\Gamma$, giving
\be
    T_{rh} = \left(90\,\Gamma^2\over 8 g_*\pi^3\, G\right)^{1/4}\,.
    \label{Trheq}
\ee

The Boltzmann equation that governs the radiation energy density $\rho_r = g_* (\pi^2/30) T^4$ during and after reheating, but before neutrino decoupling, is
\be
    {d\rho_r\over dt} + 4H\rho_r = \Gamma\rho_\Phi(t)
    \label{Boltz}
\ee
where $\Gamma$ is the rate of $\Phi\to\bar b b$, which we assume to dominate its total width, and $\rho_\Phi = \rho_0 e^{-\Gamma t}/a^3(t)$ is the energy density of $\Phi$, assumed to be nonrelativistic.  The Hubble rate is given by the Friedmann equation
\be
    H^2 \simeq {8\pi G\over 3}(\rho_\Phi + \rho_r)\,,
    \label{Friedmann}
\ee
neglecting small contributions from matter and dark energy during the BBN epoch.  The choice of $\rho_0$ is arbitrary, governing the maximum temperature of the Universe, and need only be large enough so that $T_{\rm max} > T_{rh}$, allowing the Universe to
become radiation dominated before BBN.   In practice, we choose an initial temperature $T_i = 10\,T_{rh}$, corresponding to an initial time $t_i = t_{rh}/10^4$, in accordance with the $T\sim t^{-1/4}$ scaling during reheating.  Normalizing the scale factor $a_i=1$ at $t=t_i$, one finds $a_{rh} = 10^{8/3}$, and estimating that $\rho_\Phi = \rho_r$ at $t_{rh} = 2/(3\Gamma)$, we determine
\be
    \rho_0 = e^{2/3}\pi^2 {g_*\over 30} a_{rh}^3 T_{rh}^4\,,
\ee
where $g_* = 10.75$ between the QCD phase transition and neutrino decoupling.

\begin{figure}[t!]
\centerline{\includegraphics[width=\columnwidth]{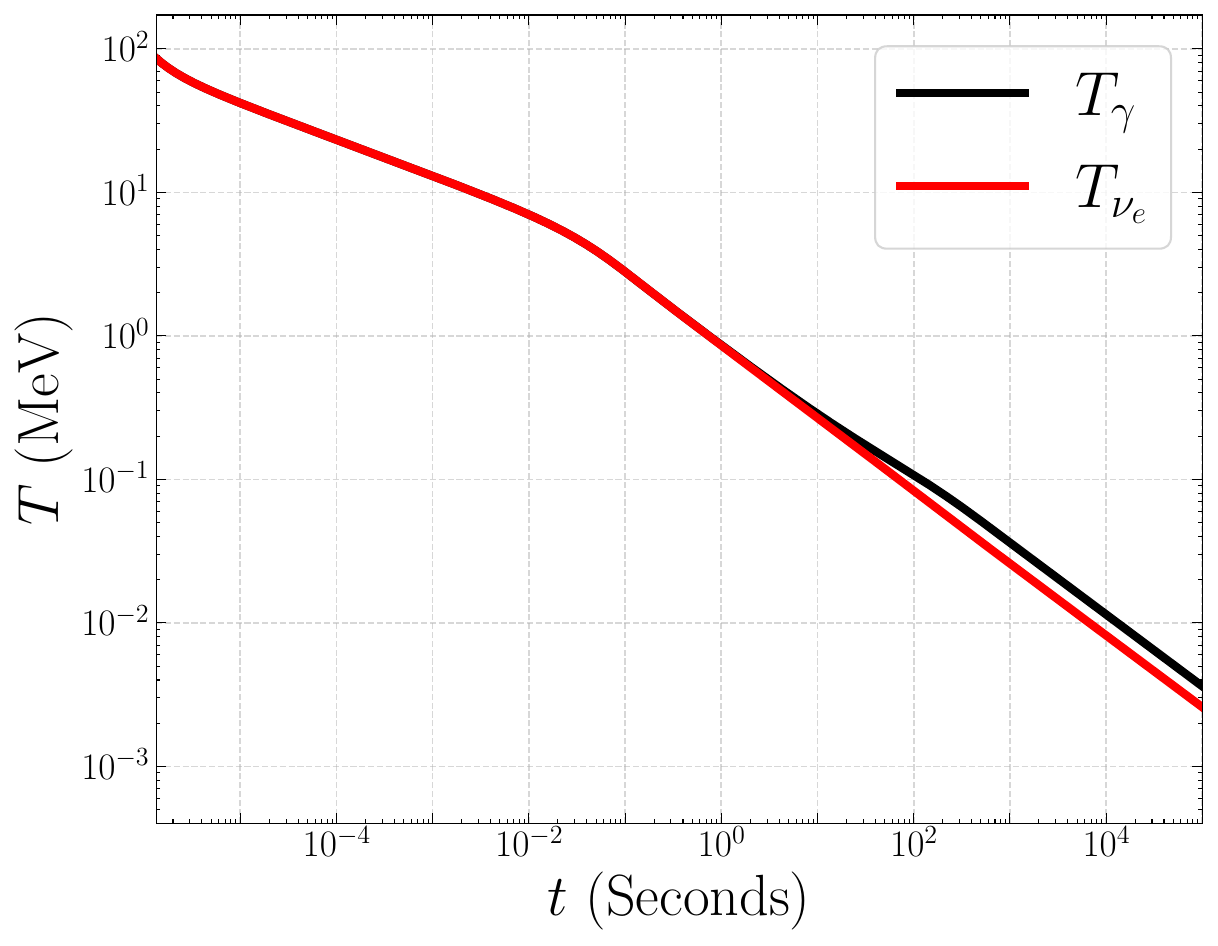}}
\caption{Time evolution of temperatures during and after reheating by $\Phi$ decays, for the example of decay rate $\Gamma = 22$/s.   Effects of $\nu$ injections by $\Phi$ decay are shown in Fig.\ \ref{fig:dTnu}.}
\label{fig:Tevol}
\end{figure}

The \texttt{BBN-simple} code \cite{Meador-Woodruff:2024due} we use for computing the background evolution uses the radiation (electrons, positrons and photons) temperature $T$ as the independent variable.  The time-dependence of $T$ in our modified scenario is implicitly determined through
\be
    {dt\over dT} = {d(\rho_e + \rho_\gamma)/dT \over -3H(\rho_e + \rho_\gamma + p_e + p_\gamma) + f\,\Gamma\rho_\Phi}\,.
    \label{dtdT}
\ee
The denominator represents $d(\rho_e+\rho_\gamma)/dt$, hence this is just the chain rule.
Below neutrino decoupling, only the fraction $f=22/43$ of decays going into $e^\pm,\gamma$ and not neutrinos affects $T$, assuming fast thermalization, while the neutrino temperatures are determined separately below. For higher temperatures, $f=1$.\footnote{More precisely, we take $f = 1 - (7/43)\sum_i\Theta_i(T)$, using Eq.\ (\ref{Theta_eq}).}
The resulting time-dependence of 
$T$ is shown as the black curve in 
Fig.\ \ref{fig:Tevol}.  It displays the expected broken power-law behavior with the break occurring at the reheat temperature $T_{rh}$. 
 One can notice that our estimate for the initial $T_i$ versus $t_i$ does not exactly match the steady-state behavior that quickly sets in afterwards; these details of the initial conditions are irrelevant to the thermal history during BBN.

\section{Neutrino temperatures}
\label{NTS}

The previous equations are not adequate to describe freeze-out of the neutrinos, when $T_\nu$ starts to fall below $T$, nor to take into account heating caused by the $\Phi$ decays.  Neglecting the latter effect, \texttt{BBN-simple} tracks $T_\nu$ according to
\be
    {dT_\nu\over dT} = -H T_\nu \left(dt\over dT\right)_{\Gamma=0}\,.
    \label{dTnudT}
\ee
In standard cosmology, this gives the expected behavior, in which $T_\nu$ tracks $T$ at temperatures above neutrino decoupling, and falls below
$T_\gamma$ when $e^\pm$ fall out of equilibrium, heating the photons but not the neutrinos, as can be seen in Fig.\ \ref{fig:Tevol}.  It is essential to use 
$(dt/dT)_0$ rather than Eq.\ (\ref{dtdT}) in Eq.\ (\ref{dTnudT}).  Using the full $dt/dT$ would give the unphysical result of $T_\nu$ exceeding $T_\gamma$ at early times before $\nu$ decoupling, as a consequence of the denominator in (\ref{dTnudT}) being too large in magnitude.

Eq.\ (\ref{dTnudT}) does not yet include the additional effects of neutrinos being injected by the $\Phi$ decays after $\nu$ decoupling.  These will increase the average neutrino energies above the standard model prediction at temperature below $\sim 1$\,MeV.  To quantify this, we take advantage of the PPPC4DMID package \cite{Cirelli:2010xx}, which has computed the showering of such particle decays into various final states, including neutrinos, as a function of 
$m_\Phi$.\footnote{The kinematics of $\Phi$ decays are identical to that of two dark matter particles, each having mass $m_\Phi/2$, annihilating from rest.}

\begin{figure}[t!]
\centerline{\includegraphics[width=\columnwidth]{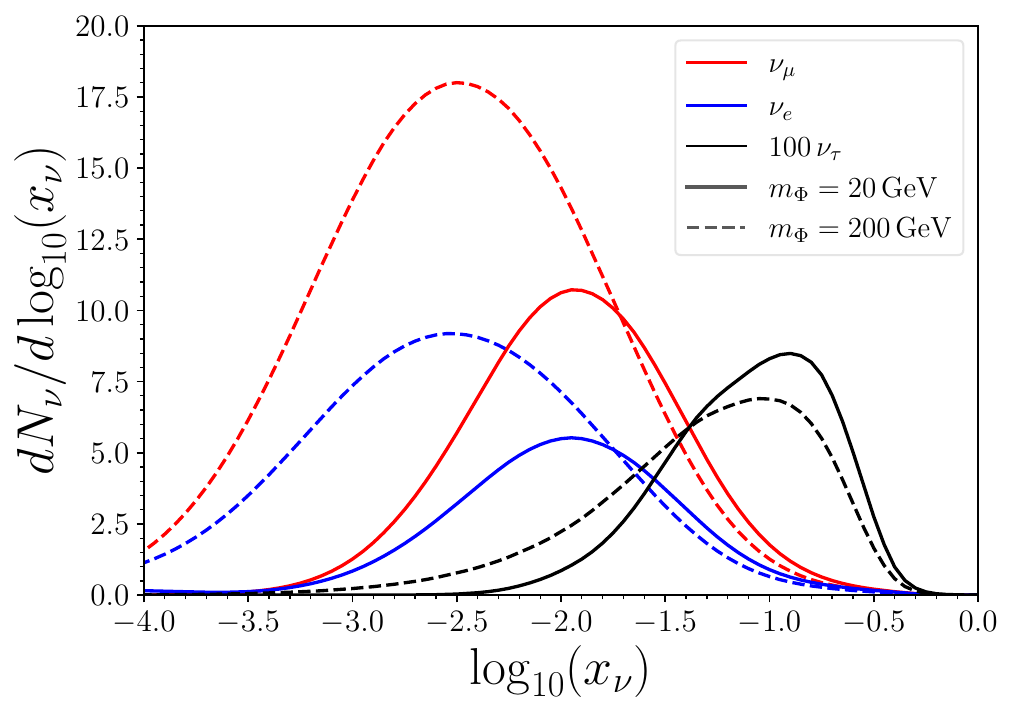}}
\caption{Differential neutrino energy spectra from the decays of $\Phi$ with masses 20 and 200\,GeV respectively, from PPPC4DMID package of Ref.\ \cite{Cirelli:2010xx}. $x$ is defined to be $x=2E/m_\Phi$, since any decay product can have a maximum energy of $m_\Phi/2$. We have enhanced the spectrum of $\nu_\tau$ by a factor 100 for better comparability.}
\label{fig:nu-spect}
\end{figure}

\begin{figure}[t!]
\centerline{\includegraphics[width=1.0\columnwidth]{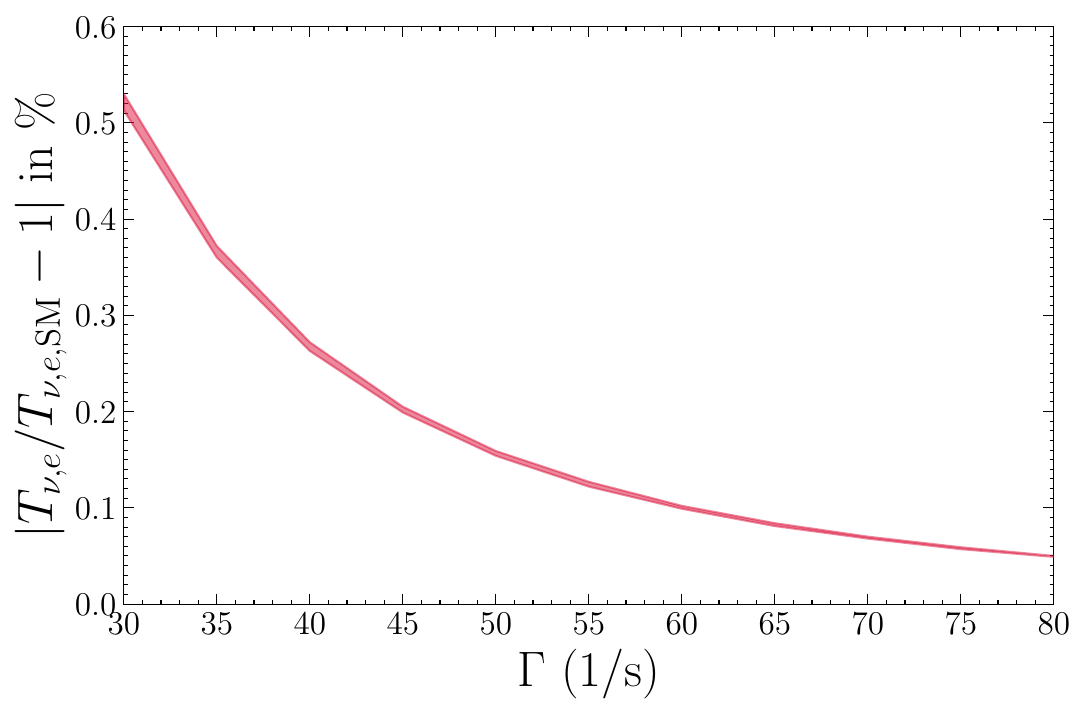}}
\caption{Percent increase in electron neutrino temperature resulting from injections of $\Phi$ decay products, as a function of $\Phi$ decay rate.  The band shows the variation with $M_\Phi \in [20,100]$\,GeV.}
\label{fig:dTnu}
\end{figure}

Examples of neutrino spectra are shown in Fig.\ \ref{fig:nu-spect}, as a function of $x=2E_\nu/m_\Phi$. $x=1$ represents the maximum energy a particle can have from the decays.  For $m_\Phi = 20\,$GeV, the integrated spectrum gives an average of 7.3 $\nu_e$'s per decay, carrying a total average energy of $\bar x = 0.16$ times $m_\Phi/2$,
$\bar E_\nu = 1.6\,$GeV.  For $m_\Phi = 200\,$GeV, the corresponding values are 16.8\,$\nu_e$'s with 
$\bar x = 0.15$, hence 15\,GeV 
of injected energy
per decay.
$\bar\nu_e$ antineutrinos will give equal contributions to those of $\nu_e$.  

Although the injected $\nu$ spectra are not thermal, for simplicity we model them as an increase in $T_\nu$ for each flavor, by equating the injected energy density with the corresponding change in a thermal distribution.
{It is known that the injection of neutrinos around the time of decoupling can lead to non-trivial effects on neutrino \emph{and} photon temperature \cite{Boyarsky:2021yoh}; however, we find that the hadronic injections following the decay of $\Phi$ have a stronger impact on BBN so that we can neglect these effects.}

Differential heating of the neutrinos by decays only starts to happen below the decoupling temperature $T_{eq,i}$ of  neutrino species $i$; otherwise the injected energy equilibrates equally between $\nu_i$ and the other radiation components.  The rate of change of $T_{\nu_i}$ coming from injections is parametrized as\footnote{The prefactor is 15/7 rather than 30/7 because $\bar x_{\nu,i}$ denotes the fraction of energy per $m_\Phi/2$, following the PPPC convention.}
\be
    \delta \dot T_{\nu_i} = \frac{15}{7}
    {\Gamma_\Phi \rho_\Phi\over \pi^2\,T_{\nu_i}^3}\,\bar x_{\nu,i}\,\Theta_i(T)\,.
\ee
Here $\bar{x}_{\nu,i}$ is the total fraction of energy relative to $M_\Phi/2$ that goes into neutrino species $\nu_i$, 
and 
$\Theta_i(T)$ models the transition for flavor $i$ to go out of equilibrium as
\be
    \Theta_i(T) = \frac12\left(1-\tanh{\ln(T/T_{eq,i})\over \Delta_\nu}\right)
    \label{Theta_eq}
\ee
with freezeout temperatures $T_{eq,e} = 1.65\,$MeV for $\nu_e$,  
$T_{eq,i} = 2.75\,$MeV for $\nu_\mu$
and $\nu_\tau$, and $\Delta_\nu=0.4$ being the logarithmic width in temperatures over which decoupling takes place.  The change in $T_{\nu,e}$ directly impacts the freezeout of $n\leftrightarrow p$ equilibrium, whereas the other flavors only affect the number of effective species $N_{\rm eff}$ influencing the Hubble rate

Some discussion is warranted about the assumptions that go into our evaluation of the temperature evolution in the electromagnetic versus neutrino sectors. Generally, the injected hadrons in the decay cascade of $\Phi$, in particular the charged ones, will lose most of their kinetic energy through efficient electromagnetic interactions, so that $\bar{x}_{\nu,i}$ {\it overestimates} the true neutrino contribution. However, the exact fraction of the neutrino energy is a complicated function of temperature and $m_\Phi$. We have thus taken a pragmatic approach: by varying the neutrino fraction between the values quoted above and a purely EM injection, we find that resulting constraints are weakened by at most $5\%$ over the mass range considered. This justifies the prescription adopted above. 

Scanning over $\Phi$ masses and lifetimes, we find a small effective increase in the electron neutrino temperature $T_{\nu_e}$ (as well as the other flavors) of order $\lesssim 0.5\,\%$, for
$\Gamma\gtrsim 30$/s.  The increase $\delta T_\nu$ depends only weakly on $M_\Phi$, as shown by red band in Fig.\ \ref{fig:dTnu}.  This reflects the fact that the integrated energies corresponding to the distributions in Fig.\ \ref{fig:nu-spect} depend only weakly on $M_\Phi$.

\begin{figure}[t!]
\centerline{\includegraphics[width=\columnwidth]{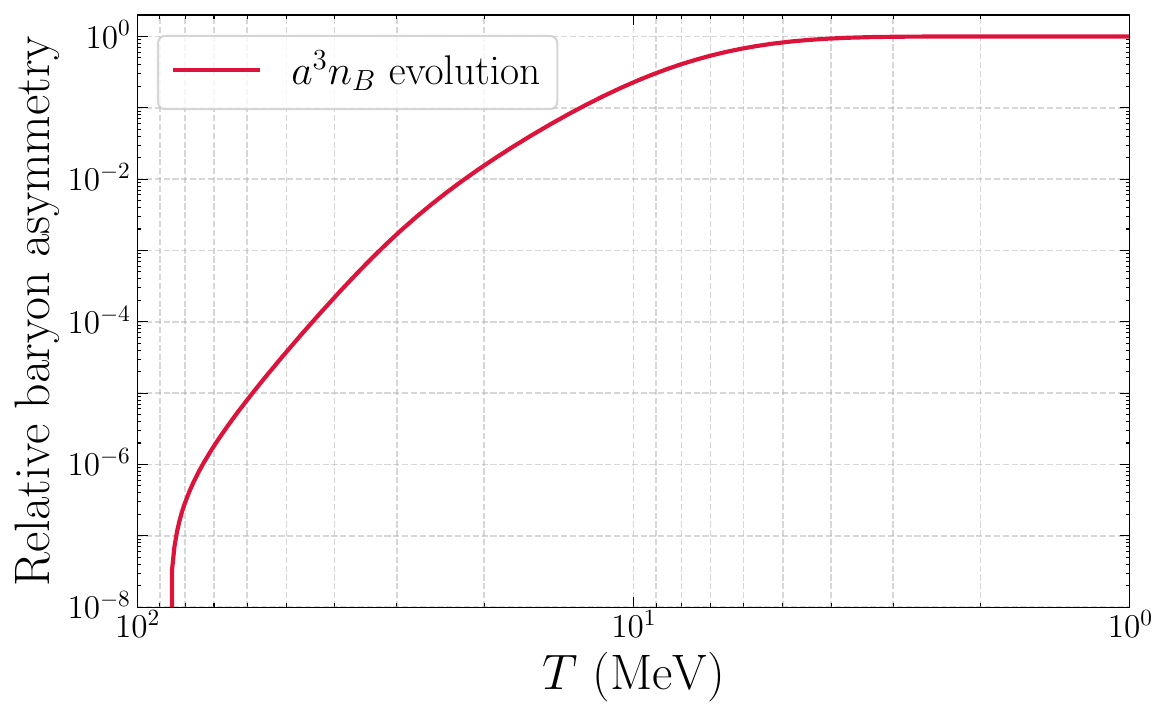}}
\centerline{\includegraphics[width=1.05\columnwidth]{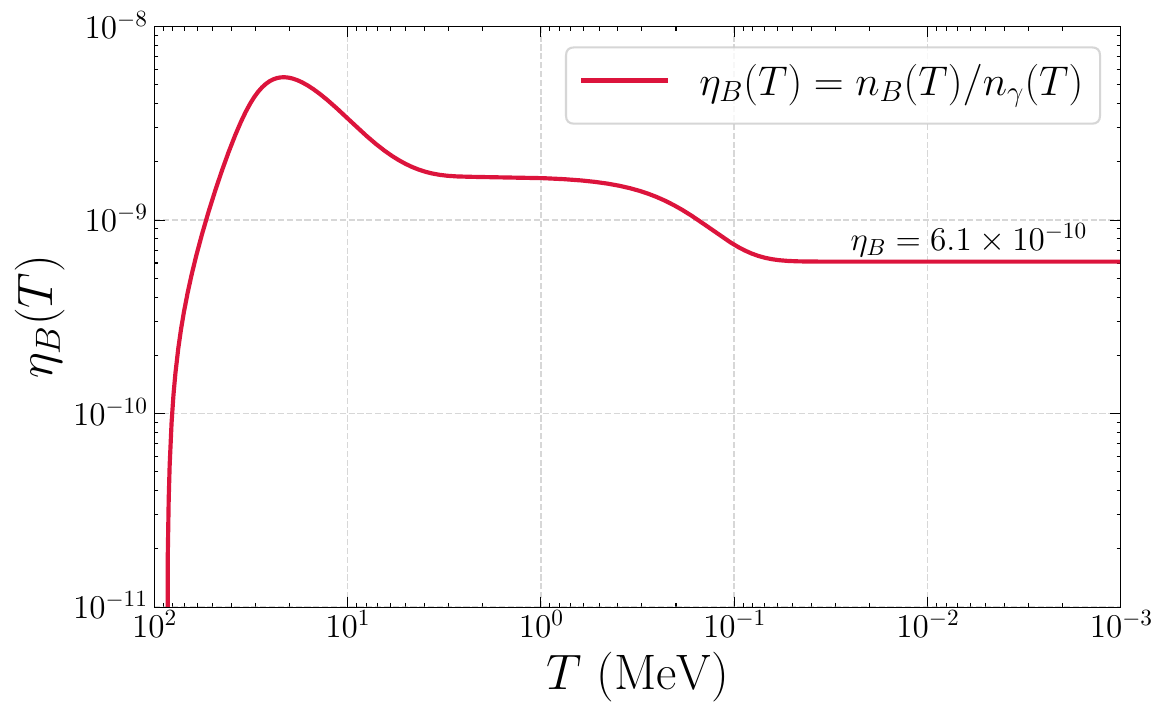}}
\caption{Temperature dependence of comoving baryon abundance (upper panel) and baryon-to-photon ratio (lower panel) resulting from residual $\Phi$ decays, for $M_\Phi = 100\,$GeV and $\Gamma_\Phi = 50$/s.}
\label{fig:etaB}
\end{figure}


\medskip
\section{Time dependence of baryon asymmetry}
\label{TDBA}

A further novel effect in $B$-mesogenesis is that the baryon-to-photon ratio $\eta_B= 6.12\times 10^{-10}$ can have residual time dependence during BBN, if the reheating temperature is sufficiently low.  This arises from the fraction of $\Phi$ particles that are still decaying.

The baryon density varies with time according to 
\be
{d\over dt}(a^3 n_B)  \sim \Gamma_B\Delta\,a^3 
\label{dnBdt}
\ee
where $a(t)$ is the scale factor and $\Delta/n_\Phi$ is the temperature-dependent net number of baryons produced per $\Phi$ decay, given in Eq.\ (25) of Ref.\ \cite{Nelson:2019fln}.\footnote{There is a factor of $M_\Phi$ missing in this equation, since $\Delta$ has units of number density.}  It is a sum over contributions from $B^0_d$ and $B^0_s$ decays:
\bea
&&    {\Delta\over n_\Phi} =
        \sum_{i=d,s} 2A_{SL,i}\Gamma_\Phi\,\tilde\Gamma_B\,|M_{12,i}|^2\nn\\
  &&     \Big[-4|M_{12,i}|^2|\Gamma_{12,i}|^2\\
        && 
              +\Gamma_B\tilde\Gamma_B(4|M_{12,i}|^2-|\Gamma_{12,i}|^2) +\Gamma_B^2\tilde\Gamma_B^2\Big]^{-1}\nn
\eea
where we have defined
\be
    \tilde\Gamma_B = \Gamma_B(1 + 2(T/T_{sc})^5),
\ee
whose second term quantifies the rate of thermal scattering, with $T_{sc} = 10.71\,$MeV.  

The terms
$M_{12,i}$ and $\Gamma_{12,i}$ are the off-diagonal elements of the $B^0_i$ mass and decay matrices, respectively, using their Standard Model values.  For simplicity, it is assumed that new physics contributions from dark decays serve only to alter the phases $\phi_{\Gamma,M}$ of $\Gamma_{12,i}$ and $M_{12,i}$, and furthermore that these phases remain small, albeit possibly larger than their SM values.\footnote{This must be the case for $A_{SL,s}$ in order to generate a large enough baryon asymmetry, but not necessarily for $A_{SL,d}$.}  The new physics contributions are encoded in the
parameters $A_{SL,i} = |\Gamma_{12,i}/M_{12,i}|\sin(\phi_\Gamma - \phi_M)$, whose experimentally allowed ranges are larger than their predicted SM values.  

Typically $B^0_s$ decays dominate the production of the baryon asymmetry, with only small contributions from $B_d^0$, unless the reheat temperatures is below 10\,MeV.
For definiteness, we take $A_{SL,i}$ to be at the top of their allowed ranges ($-4\times 10^{-4}$ for $B_d^0$ and
$8.8\times 10^{-4}$ for $B_s^0$), to fix their relative contributions.  Their magnitudes are then rescaled  (corresponding to the overall proportionality coefficient 
in Eq.\ (\ref{dnBdt})) by fixing the late-time baryon-to-photon ratio as 
$\eta = 6.15\times 10^{-10}$, consistent with Planck \cite{Planck:2018vyg}.\footnote{We have made sure that within the parameter space we are probing there is sufficient baryogenesis to explain the observed value. The rescaling then just corresponds to varying the $A_{SL,i}$ within their allowed ranges.} 
Numerically, our results can be well approximated by ignoring the $B_d^0$ contribution.

As an example, the temperature dependence of comoving baryon number $(a^3 n_B)$  and  baryon-to-photon ratio $\eta_B = n_B/n_\gamma$ are shown in Fig.\ \ref{fig:etaB}, for $M_\Phi = 100\,$GeV and $\Gamma_\Phi = 50$\,s$^{-1}$. $a^3n_B$ increases monotonically during reheating, after which further production is exponentially suppressed. On the other hand, $\eta_B$ first reaches a maximum and then decreases from entropy production during reheating. At $T\sim m_e$, photon heating from $e^+e^-$ annihilation causes a further decrease in $\eta_B$.

\begin{figure}[t!]
\centerline{\includegraphics[width=\columnwidth]{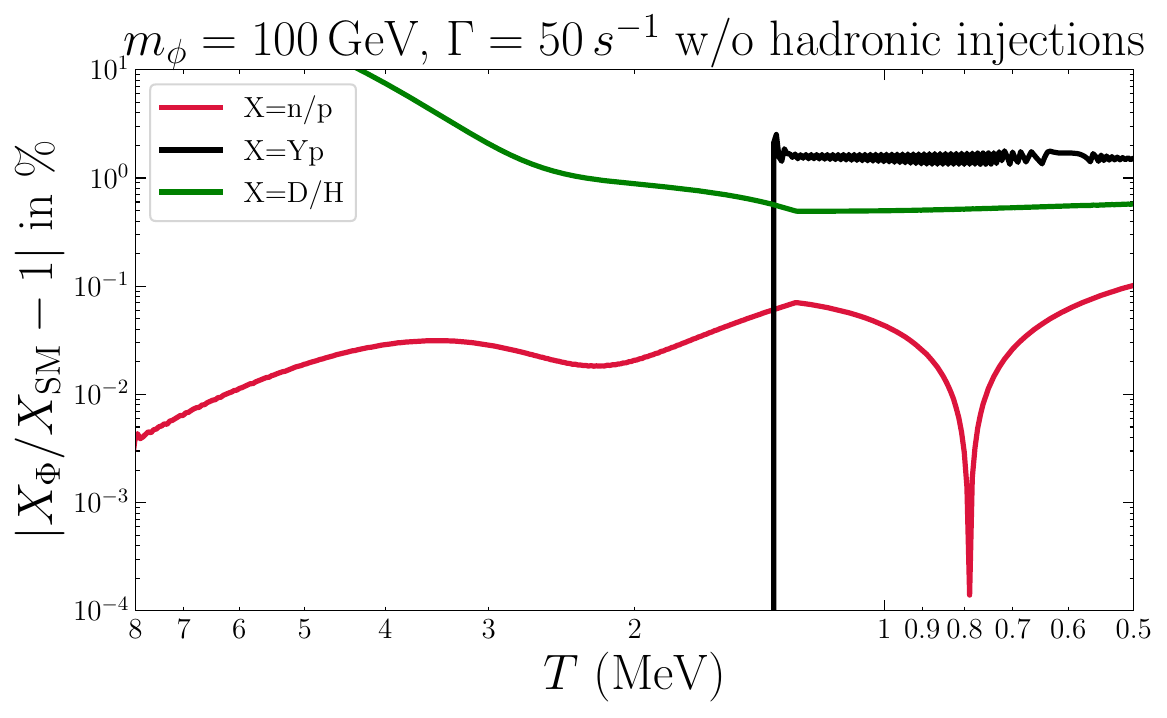}}
\caption{Percent difference in $n/p$, $Y_p$ and the deuterium abundance $D/H$ as a function of $T$, from effects of $\Phi$ decays excluding hadronic injections.  The example of $M_\Phi = 100\,$GeV and $\Gamma = 50\,$s$^{-1}$ is illustrated.}
\label{fig:resid1}
\end{figure}

The effects of neutrino heating and time-dependence of $\eta_B$ on BBN are illustrated in Fig.\ \ref{fig:resid1} for the case of $M_\Phi = 100\,$GeV and $\Gamma = 50\,$s$^{-1}$.  The dips in the curves correspond to the instantaneous values passing through their standard model value ({\it e.g.}, from below to above).
Larger deviations are induced by hadronic injections, as we discuss next. 

\section{Hadronic Injection effects on BBN}
\label{HIE}
The remaining nonstandard feature is that residual $B$ decays inject energetic hadrons
($p$, $n$, $\pi^\pm$, $K^\pm$, $K_L$) into the plasma, that can skew the neutron-proton equilibrium during BBN.  Any such perturbation has a direct impact on helium production.  To accurately assess their effect, one must simulate the hadronic showering of $b$ quark decays using a Monte Carlo package like \texttt{PYTHIA}. 

These effects are outside of the scope of \texttt{BBN-simple}, which we used so far to compute the modified expansion rate and temperatures, but they are included in the code \texttt{AlterAlterBBN}
(\texttt{AABBN}) \cite{Hufnagel:2017dgo, Bianco:2025boy, Bianco:2026dvc} (based on \texttt{AlterBBN} \cite{Arbey:2011nf,Arbey:2018zfh}).  We therefore use the modified cosmology from \texttt{BBN-simple} as input to \texttt{AABBN} to account for the hadronic injections.  

It is well-known that photo- and hadro-dissociation of the light element can lead to strong constraints on hadron and photon injection in the early Universe. However, they are negligible effects in the present context:  Ref.\ \cite{Bianco:2025boy} found that interconversion bounds are dominant up to lifetimes of
10-100\,s, relevant for $B$-mesogenesis, and only for longer lived $\Phi$'s,  dissociation becomes important. This is because only for 
temperatures below $\sim 100\,$keV is the stopping power of the plasma sufficiently weak to allow for
energetic neutrons that can dissociate nuclei. We explicitly checked that exclusion of this effect has no impact on our results.
 
\begin{figure}[t!]
\centerline{\includegraphics[width=\columnwidth]{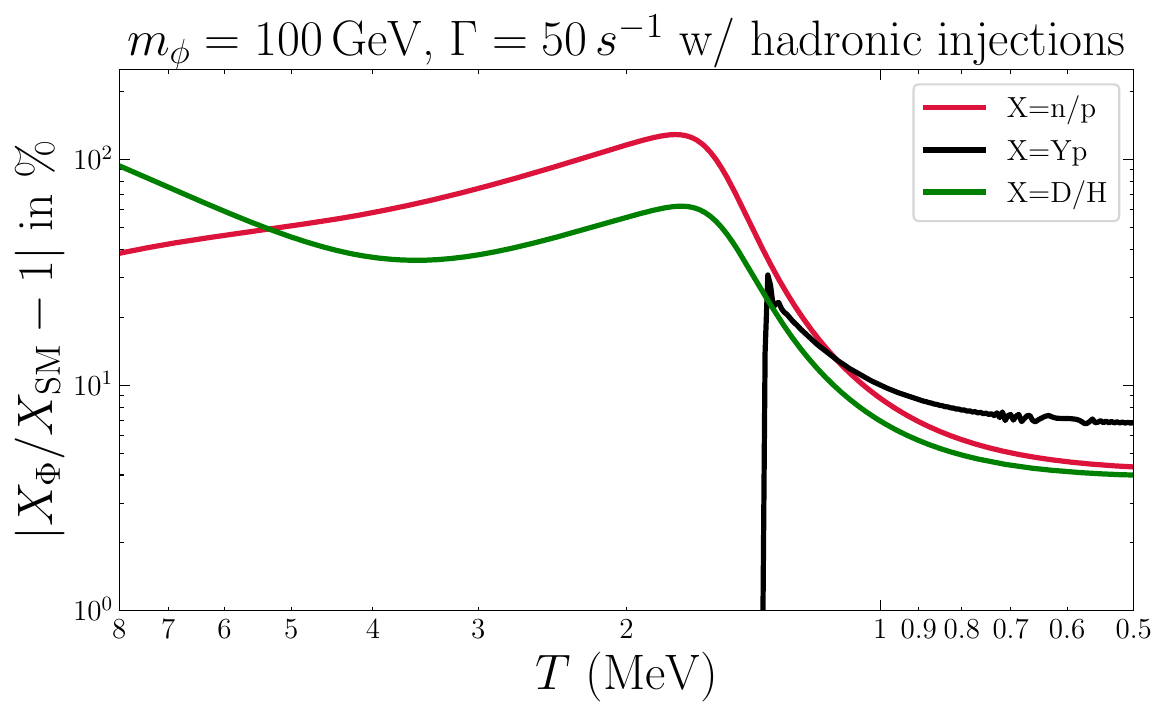}}
\caption{Like Fig.\ \ref{fig:resid1}, but including effects of hadronic injections on $n/p$ interconversion.}
\label{fig:resid2}
\end{figure}

In Fig.\ \ref{fig:resid2} we show the temperature 
evolution of $n/p$, $Y_p$ and $D/H$, including hadronic injection effects, at the same $\Phi$ mass and lifetime as in Fig.\ \ref{fig:resid1}.  The deviation from the standard model increases by a factor of $\sim 10$ compared to including neutrino heating and time-dependence of $\eta_B$. 

\begin{figure}[t!]
\centerline{\includegraphics[width=\columnwidth]{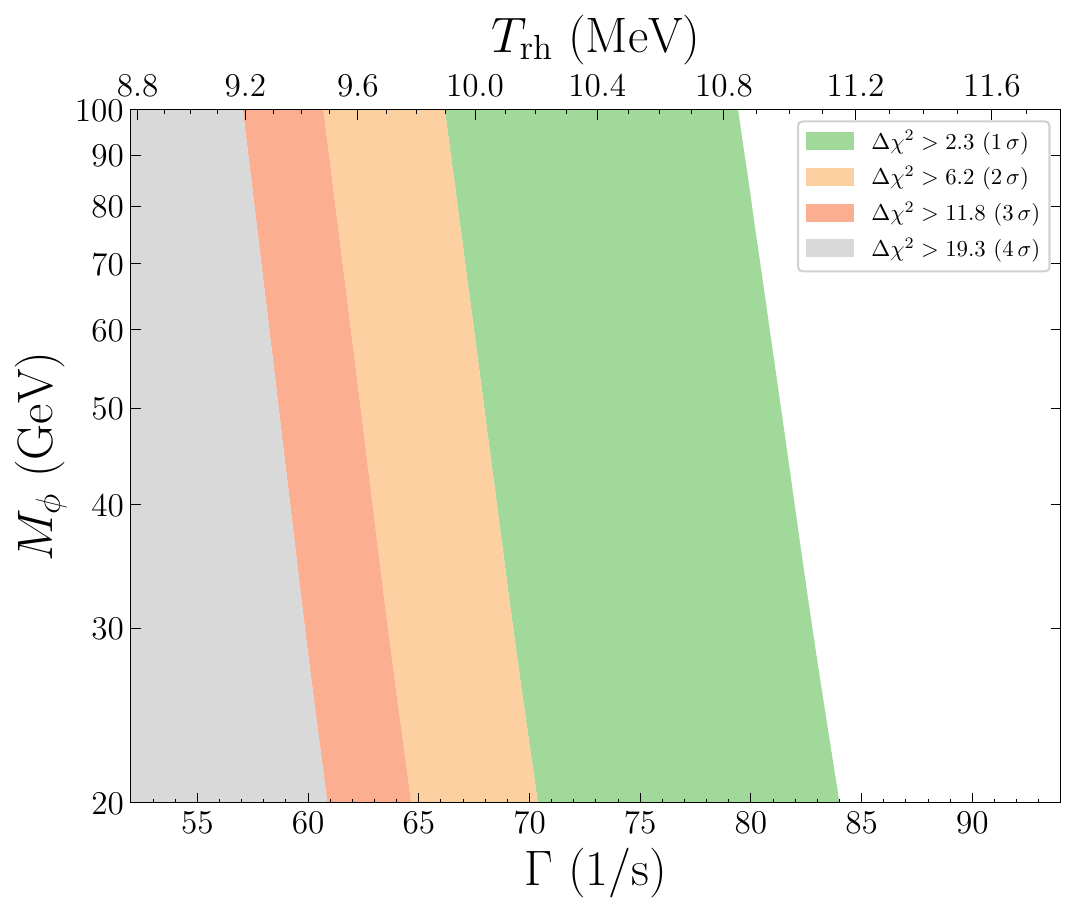}}
\caption{Limits on the decay rate/reheating temperature from BBN.  Regions toward left (orange and grey) are excluded.}
\label{fig:contours}
\end{figure}

\section{Exclusion regions}
\label{XRS}
Using the methodology described above, we have explored the [$M_\Phi$, $\Gamma$] parameter space of $B$-mesogenesis to determine the regions allowed by BBN.  The long-standing $^7$Li problem is not ameliorated by the new physics, and so we neglect this contribution as is customary.  Our likelihood  estimate therefore only incorporates $Y_p$ and $D/H$, using the mean values and uncertainties \cite{ParticleDataGroup:2026mpi, Yeh:2026pil}
\bea
    Y_p &=& 0.2458\pm 0.0013\,,\nn\\
    D/H &=& (2.508\pm 0.029)
\times 10^{-5}\,.
\eea
We further incorporate the systematic uncertainties arising from \texttt{AABBN} which includes the uncertainties of the fusion rates but in particular an estimated uncertainty on the interconversion rates which dominates the final error budget. For more information on the error propagation in \texttt{AABBN} we refer to Ref.~\cite{Bianco:2026dvc}. The inclusion of these uncertainties loosens the constraints mildly.

Scanning over the mass and decay rate of $\Phi$, this likelihood yields the confidence intervals  shown in Fig.\ \ref{fig:contours}.  The top $x$-axis shows the correspondence to the reheating temperature using Eq.\ (\ref{Trheq}).  At $M_\Phi = 100$\,GeV, the 3\,$\sigma$ limit is $T_{rh}\gtrsim 9.5\,$MeV, while at $M_\Phi = 20$\,GeV, it is slightly higher, $T_{rh}\gtrsim 9.8\,$MeV.
The dependence  can be fit as
\be
    T_{rh} > \left(10.3 - 0.42\log_{10}{M_\Phi\over{\rm GeV}}\right)\,{\rm MeV}\,.
    \label{Trhbound}
\ee
Numerically, this behavior continues to arbitrarily large values of $M_\Phi$, but $B$-mesogenesis is only able to produce a large enough baryon asymmetry for $M_\Phi\lesssim 100$\,GeV.  Qualitatively, one can understand that larger $M_\Phi$ leads to small number densities $n_\Phi$, hence fewer hadronic injections,
and moreover, the number of injected hadrons scales more slowly than linearly with $M_\Phi$.
Although the injected hadrons have higher energy as $M_\Phi$ increases, 
they quickly thermalize; hence the lowering of the number density is not fully compensated by the increase in energy. 

\section{Summary}
\label{SumS}
$B$-mesogenesis is a highly predictive model of baryogenesis, testable at low energies. It requires a low reheating temperature $T_{rh}\lesssim 140\,$MeV, following a late period of matter domination by a heavy scalar $\Phi$ that matter dominates the Universe at late times, until it decays into $b$ quarks.  

This work is the first to quantitatively answer the question of how low can $T_{rh}$ be, which depends on the decay rate of $\Phi$.  We have shown that Big Bang Nucleosynthesis requires $T_{rh}\gtrsim 9.5\,$MeV  if $M_\Phi \lesssim 100\,$GeV, corresponding to a $\Phi$ lifetime
$\tau \lesssim 0.02\,$s.  These bounds depend weakly upon $M_\Phi$, as given in Eq.\ (\ref{Trhbound}).  Significant relaxation of the limit could occur for very large $M_\Phi\gtrsim 10^5$\,GeV, for example, but such values would not be compatible with a large enough baryon asymmetry, which scales as
$1/M_\Phi$.

In principle, the limits also depend upon 
the relative values of the semileptonic CP decay asymmetries $A_{SL,d}$ and $A_{SL,b}$ of $B^0_d$ and $B^0_s$, but numerically we find that they are insensitive to this uncertainty.  Throughout this work we have assumed that the baryon asymmetry is dominated by the contributions of neutral $B$ meson decays.  However since the limits depend mainly on the rate of $B$ decays, we do not expect the results to differ significantly for the scenario of charged $B$-mesogenesis \cite{Elahi:2021jia,BurgosMarcos:2026lwv}, in which the baryon asymmetry arises from charged rather than neutral meson decays.

\bigskip
{\bf Acknowledgments.}  We thank Jasmine Boyd, Nicolas Montano Grajales and John Wolfart for their collaboration during the initial phase of this work, 
and 
Gonzalo Alonso-Alvarez, Marco Cirelli,
Miguel Escudero, Aidan Meador-Woodruff, Dragan Huterer, Huangyu Xiao, and Kai Schmidt-Hoberg for helpful correspondence.  
JC thanks the Niels Bohr International Academy for its hospitality during the completion of this work.
JC and YX are supported by the Natural Sciences and Engineering Research Council (NSERC) of Canada.
JF is supported by an ERC StG grant (“AstroDarkLS”, Grant No. 101117510).

\bibliographystyle{utphys}
\bibliography{sample}

@article{Meador-Woodruff:2024due,
    author = "Meador-Woodruff, Aidan and Huterer, Dragan",
    title = "{BBN-simple: How to bake a universe-sized cake}",
    eprint = "2412.07893",
    archivePrefix = "arXiv",
    primaryClass = "astro-ph.CO",
    doi = "10.1016/j.newar.2025.101722",
    journal = "New Astron. Rev.",
    volume = "100",
    pages = "101722",
    year = "2025"
}

@article{Elahi:2021jia,
    author = "Elahi, Fatemeh and Elor, Gilly and McGehee, Robert",
    title = "{Charged B mesogenesis}",
    eprint = "2109.09751",
    archivePrefix = "arXiv",
    primaryClass = "hep-ph",
    reportNumber = "LCTP-21-24, MITP-21-041",
    doi = "10.1103/PhysRevD.105.055024",
    journal = "Phys. Rev. D",
    volume = "105",
    number = "5",
    pages = "055024",
    year = "2022"
}

@article{BurgosMarcos:2026lwv,
    author = "Burgos Marcos, M. and Verheyden, A. and Vos, K. K.",
    title = "{From B$_{c}$ mesons to the baryon asymmetry: a unified B mesogenesis framework}",
    eprint = "2601.19651",
    archivePrefix = "arXiv",
    primaryClass = "hep-ph",
    reportNumber = "Nikhef 2026-001",
    doi = "10.1007/JHEP09(2026)077",
    journal = "JHEP",
    volume = "09",
    pages = "077",
    year = "2026"
}

@article{Cirelli:2010xx,
    author = "Cirelli, Marco and Corcella, Gennaro and Hektor, Andi and Hutsi, Gert and Kadastik, Mario and Panci, Paolo and Raidal, Martti and Sala, Filippo and Strumia, Alessandro",
    title = "{PPPC 4 DM ID: A Poor Particle Physicist Cookbook for Dark Matter Indirect Detection}",
    eprint = "1012.4515",
    archivePrefix = "arXiv",
    primaryClass = "hep-ph",
    reportNumber = "CERN-PH-TH-2010-057, SACLAY-T10-025, IFUP-TH-2010-44",
    doi = "10.1088/1475-7516/2012/10/E01",
    journal = "JCAP",
    volume = "03",
    pages = "051",
    year = "2011",
    note = "[Erratum: JCAP 10, E01 (2012)]"
}

@article{Arbey:2018zfh,
    author = "Arbey, A. and Auffinger, J. and Hickerson, K. P. and Jenssen, E. S.",
    title = "{AlterBBN v2: A public code for calculating Big-Bang nucleosynthesis constraints in alternative cosmologies}",
    eprint = "1806.11095",
    archivePrefix = "arXiv",
    primaryClass = "astro-ph.CO",
    reportNumber = "CERN-TH-2018-146",
    doi = "10.1016/j.cpc.2019.106982",
    journal = "Comput. Phys. Commun.",
    volume = "248",
    pages = "106982",
    year = "2020"
}

@article{Elor:2018twp,
    author = "Elor, Gilly and Escudero, Miguel and Nelson, Ann",
    title = "{Baryogenesis and Dark Matter from $B$ Mesons}",
    eprint = "1810.00880",
    archivePrefix = "arXiv",
    primaryClass = "hep-ph",
    reportNumber = "KCL-18-53, IFIC-18-35",
    doi = "10.1103/PhysRevD.99.035031",
    journal = "Phys. Rev. D",
    volume = "99",
    number = "3",
    pages = "035031",
    year = "2019"
}

@article{deSalas:2015glj,
    author = "de Salas, P. F. and Lattanzi, M. and Mangano, G. and Miele, G. and Pastor, S. and Pisanti, O.",
    title = "{Bounds on very low reheating scenarios after Planck}",
    eprint = "1511.00672",
    archivePrefix = "arXiv",
    primaryClass = "astro-ph.CO",
    reportNumber = "IFIC-15-70",
    doi = "10.1103/PhysRevD.92.123534",
    journal = "Phys. Rev. D",
    volume = "92",
    number = "12",
    pages = "123534",
    year = "2015"
}

@article{Planck:2018vyg,
    author = "Aghanim, N. and others",
    collaboration = "Planck",
    title = "{Planck 2018 results. VI. Cosmological parameters}",
    eprint = "1807.06209",
    archivePrefix = "arXiv",
    primaryClass = "astro-ph.CO",
    doi = "10.1051/0004-6361/201833910",
    journal = "Astron. Astrophys.",
    volume = "641",
    pages = "A6",
    year = "2020",
    note = "[Erratum: Astron.Astrophys. 652, C4 (2021)]"
}

@article{Sakharov:1967dj,
    author = "Sakharov, A. D.",
    title = "{Violation of CP Invariance, C asymmetry, and baryon asymmetry of the universe}",
    doi = "10.1070/PU1991v034n05ABEH002497",
    journal = "Pisma Zh. Eksp. Teor. Fiz.",
    volume = "5",
    pages = "32--35",
    year = "1967"
}

@article{Bianco:2025boy,
    author = "Bianco, Sara and Depta, Paul Frederik and Frerick, Jonas and Hambye, Thomas and Hufnagel, Marco and Schmidt-Hoberg, Kai",
    title = "{Photo- and hadrodisintegration constraints on massive relics decaying into neutrinos}",
    eprint = "2505.01492",
    archivePrefix = "arXiv",
    primaryClass = "hep-ph",
    reportNumber = "DESY-25-056, ULB-TH/25-03",
    doi = "10.1088/1475-7516/2025/11/072",
    journal = "JCAP",
    volume = "11",
    pages = "072",
    year = "2025"
}

@article{Bianco:2026dvc,
    author = "Bianco, Sara and Frerick, Jonas and Hufnagel, Marco and Schmidt-Hoberg, Kai",
    title = "{Improved Big Bang Nucleosynthesis constraints on decaying massive relics}",
    eprint = "2605.26824",
    archivePrefix = "arXiv",
    primaryClass = "hep-ph",
    reportNumber = "ULB-TH/26-21, DESY-26-066",
    month = "5",
    year = "2026"
}

@article{Farrar:1993sp,
    author = "Farrar, Glennys R. and Shaposhnikov, M. E.",
    title = "{Baryon asymmetry of the universe in the minimal Standard Model}",
    eprint = "hep-ph/9305274",
    archivePrefix = "arXiv",
    reportNumber = "RU-93-10, CERN-TH-6729-92",
    doi = "10.1103/PhysRevLett.70.2833",
    journal = "Phys. Rev. Lett.",
    volume = "70",
    pages = "2833--2836",
    year = "1993",
    note = "[Erratum: Phys.Rev.Lett. 71, 210 (1993)]"
}

@article{Farrar:1993hn,
    author = "Farrar, Glennys R. and Shaposhnikov, M. E.",
    title = "{Baryon asymmetry of the universe in the standard electroweak theory}",
    eprint = "hep-ph/9305275",
    archivePrefix = "arXiv",
    reportNumber = "CERN-TH-6732-93, CERN-TH-6734-93, RU-93-11",
    doi = "10.1103/PhysRevD.50.774",
    journal = "Phys. Rev. D",
    volume = "50",
    pages = "774",
    year = "1994"
}

@article{Fukugita:1986hr,
    author = "Fukugita, M. and Yanagida, T.",
    title = "{Baryogenesis Without Grand Unification}",
    reportNumber = "RIFP-641",
    doi = "10.1016/0370-2693(86)91126-3",
    journal = "Phys. Lett. B",
    volume = "174",
    pages = "45--47",
    year = "1986"
}

@article{Davidson:2008bu,
    author = "Davidson, Sacha and Nardi, Enrico and Nir, Yosef",
    title = "{Leptogenesis}",
    eprint = "0802.2962",
    archivePrefix = "arXiv",
    primaryClass = "hep-ph",
    doi = "10.1016/j.physrep.2008.06.002",
    journal = "Phys. Rept.",
    volume = "466",
    pages = "105--177",
    year = "2008"
}

@article{Barni:2025ifb,
    author = "Barni, Giulio",
    title = "{Electroweak Baryogenesis with BARYONET: a self-contained review of the WKB approach}",
    eprint = "2510.21915",
    archivePrefix = "arXiv",
    primaryClass = "hep-ph",
    month = "10",
    year = "2025"
}

@article{Cohen:1993nk,
    author = "Cohen, Andrew G. and Kaplan, D. B. and Nelson, A. E.",
    title = "{Progress in electroweak baryogenesis}",
    eprint = "hep-ph/9302210",
    archivePrefix = "arXiv",
    reportNumber = "UCSD-PTH-93-02, BUHEP-93-4",
    doi = "10.1146/annurev.ns.43.120193.000331",
    journal = "Ann. Rev. Nucl. Part. Sci.",
    volume = "43",
    pages = "27--70",
    year = "1993"
}

@article{Gavela:1994dt,
    author = "Gavela, M. B. and Hernandez, P. and Orloff, J. and Pene, O. and Quimbay, C.",
    title = "{Standard model CP violation and baryon asymmetry. Part 2: Finite temperature}",
    eprint = "hep-ph/9406289",
    archivePrefix = "arXiv",
    reportNumber = "CERN-TH-7263-94, LPTHE-ORSAY-94-49, HUTP-94-A015, HD-THEP-94-20, FTUAM-94-14, NSF-ITP-94-65",
    doi = "10.1016/0550-3213(94)00410-2",
    journal = "Nucl. Phys. B",
    volume = "430",
    pages = "382--426",
    year = "1994"
}

@article{Gavela:1993ts,
    author = "Gavela, M. B. and Hernandez, P. and Orloff, J. and Pene, O.",
    title = "{Standard model CP violation and baryon asymmetry}",
    eprint = "hep-ph/9312215",
    archivePrefix = "arXiv",
    reportNumber = "CERN-TH-7081-93, LPTHE-ORSAY-93-48, HUTP-93-A036, HD-THEP-93-45",
    doi = "10.1142/S0217732394000629",
    journal = "Mod. Phys. Lett. A",
    volume = "9",
    pages = "795--810",
    year = "1994"
}

@article{Alonso-Alvarez:2021qfd,
    author = "Alonso-{\'A}lvarez, Gonzalo and Elor, Gilly and Escudero, Miguel",
    title = "{Collider signals of baryogenesis and dark matter from B mesons: A roadmap to discovery}",
    eprint = "2101.02706",
    archivePrefix = "arXiv",
    primaryClass = "hep-ph",
    reportNumber = "TUM-HEP 1299/20",
    doi = "10.1103/PhysRevD.104.035028",
    journal = "Phys. Rev. D",
    volume = "104",
    number = "3",
    pages = "035028",
    year = "2021"
}

@article{Nelson:2019fln,
    author = "Nelson, Ann E. and Xiao, Huangyu",
    title = "{Baryogenesis from B Meson Oscillations}",
    eprint = "1901.08141",
    archivePrefix = "arXiv",
    primaryClass = "hep-ph",
    doi = "10.1103/PhysRevD.100.075002",
    journal = "Phys. Rev. D",
    volume = "100",
    number = "7",
    pages = "075002",
    year = "2019"
}

@article{Kawasaki:2000en,
    author = "Kawasaki, M. and Kohri, Kazunori and Sugiyama, Naoshi",
    title = "{MeV scale reheating temperature and thermalization of neutrino background}",
    eprint = "astro-ph/0002127",
    archivePrefix = "arXiv",
    doi = "10.1103/PhysRevD.62.023506",
    journal = "Phys. Rev. D",
    volume = "62",
    pages = "023506",
    year = "2000"
}

@article{Hannestad:2004px,
    author = "Hannestad, Steen",
    title = "{What is the lowest possible reheating temperature?}",
    eprint = "astro-ph/0403291",
    archivePrefix = "arXiv",
    doi = "10.1103/PhysRevD.70.043506",
    journal = "Phys. Rev. D",
    volume = "70",
    pages = "043506",
    year = "2004"
}

@article{Boyarsky:2021yoh,
    author = "Boyarsky, Alexey and Ovchynnikov, Maksym and Sabti, Nashwan and Syvolap, Vsevolod",
    title = "{When feebly interacting massive particles decay into neutrinos: The Neff story}",
    eprint = "2103.09831",
    archivePrefix = "arXiv",
    primaryClass = "hep-ph",
    reportNumber = "KCL-2021-05",
    doi = "10.1103/PhysRevD.104.035006",
    journal = "Phys. Rev. D",
    volume = "104",
    number = "3",
    pages = "035006",
    year = "2021"
}

@article{Hufnagel:2017dgo,
    author = "Hufnagel, Marco and Schmidt-Hoberg, Kai and Wild, Sebastian",
    title = "{BBN constraints on MeV-scale dark sectors. Part I. Sterile decays}",
    eprint = "1712.03972",
    archivePrefix = "arXiv",
    primaryClass = "hep-ph",
    reportNumber = "DESY-17-211",
    doi = "10.1088/1475-7516/2018/02/044",
    journal = "JCAP",
    volume = "02",
    pages = "044",
    year = "2018"
}

@article{Arbey:2011nf,
    author = "Arbey, Alexandre",
    title = "{AlterBBN: A program for calculating the BBN abundances of the elements in alternative cosmologies}",
    eprint = "1106.1363",
    archivePrefix = "arXiv",
    primaryClass = "astro-ph.CO",
    reportNumber = "CERN-PH-TH-2011-135, LYCEN-2011-06",
    doi = "10.1016/j.cpc.2012.03.018",
    journal = "Comput. Phys. Commun.",
    volume = "183",
    pages = "1822--1831",
    year = "2012"
}

@article{ParticleDataGroup:2026mpi,
    author = "Takahashi, F. and others",
    collaboration = "Particle Data Group",
    title = "{Review of Particle Physics}",
    doi = "10.1142/s0217751x26300115",
    journal = "Int. J. Mod. Phys. A",
    volume = "41",
    number = "22",
    pages = "2630011",
    year = "2026"
}

@article{Giudice:2000ex,
    author = "Giudice, Gian Francesco and Kolb, Edward W. and Riotto, Antonio",
    title = "{Largest temperature of the radiation era and its cosmological implications}",
    eprint = "hep-ph/0005123",
    archivePrefix = "arXiv",
    reportNumber = "SNS-PH-00-05, FERMILAB-PUB-00-075-A, CERN-TH-2000-107",
    doi = "10.1103/PhysRevD.64.023508",
    journal = "Phys. Rev. D",
    volume = "64",
    pages = "023508",
    year = "2001"
}

@article{Yeh:2026pil,
    author = "Yeh, Tsung-Han and Olive, Keith A. and Fields, Brian D. and Aver, Erik and Pogge, Richard W. and Rogers, Noah S. J. and Skillman, Evan D. and Weller, Miqaela K.",
    title = "{The LBT Y$_{p}$ Project. V. Cosmological Implications of a New Determination of Primordial $^{4}$He}",
    eprint = "2601.22239",
    archivePrefix = "arXiv",
    primaryClass = "astro-ph.CO",
    reportNumber = "FTPI-MINN-26-03, UMN-TH-4519/26",
    doi = "10.3847/1538-4357/ae879c",
    journal = "Astrophys. J.",
    volume = "1008",
    number = "2",
    pages = "241",
    year = "2026"
}

\end{document}